\documentclass[runningheads]{llncs}
\usepackage{graphicx}
\begin{document}
\title{Towards a secure behavior modeling for IoT networks using Blockchain}
% A Framework for Secure Behavior Monitoring for IoT Environments using Blockchain
\titlerunning{Secure behavior modeling for IoT networks}
%%%Behavioral profiling framework for IoT Environments in a Blockchain infrastructure
%
\author{Jawad Ali\inst{1} %%\orcidID{0000-0002-6015-0663} <-- Jawad ORCID
	\and
Ahmad Sharafidz Khalid\inst{1}
 \and
Eiad Yafi\inst{1} 
\and
Shahrulniza Musa\inst{1}
\and
Waqas Ahmed\inst{2}
}
\authorrunning{Jawad et al.}

\institute{
Malaysian Institute of Information Technology, \\ Universiti 
	Kuala Lumpur, Malaysia.\\
\email{jawad.ali@s.unikl.edu.my,
ahmads@unikl.edu.my, \\
eiad@unikl.edu.my, shahrulniza@unikl.edu.my,
}
\and
UniKL Business School, Universiti Kuala Lumpur, Malaysia.
\email{waqas.ahmed@s.unikl.edu.my}
}
\maketitle              % typeset the header of the contribution
%
%UniKL Business School, Universiti Kuala Lumpur, Malaysia
%%%%
\begin{abstract}
Internet of Things (IoT) occupies a vital aspect of our everyday lives. IoT networks composed of smart-devices which communicate and transfer the information without the physical intervention of humans. Due to such proliferation and autonomous nature of IoT systems make these devices threatened and prone to a severe kind of threats. In this paper, we introduces a behavior capturing, and verification procedures in blockchain supported smart-IoT systems that can be able to show the trust-level confidence to outside networks. We defined a custom \emph{Behavior Monitor} and implement on a selected node that can extract the activity of each device and analyzes the behavior using deep machine learning strategy. Besides, we deploy Trusted Execution Technology (TEE) which can be used to provide a secure execution environment (enclave) for sensitive application code and data on the blockchain. Finally, in the evaluation phase we analyze various IoT devices data that is infected by Mirai attack. The evaluation results show the strength of our proposed method in terms of accuracy and time required for detection.
%\end{abstract}

\end{abstract}
\section{Introduction}

Currently, in the modern world Internet of Things (IoT) is swiftly growing and involved in every aspect of our daily computations. According to the industry-leading experts' argument that more than 50 billion of IoT devices will be deployed by 2020 \cite{gartner}. Things in IoT are the collection of web-enabled devices that use embedded processors, sensors, micro-controllers and communication hardware (exchange of data from different environments). Such rich communication in IoT devices generates an enormous dataset which in turn to use for various dependent services.

Apart from this, IoT allows the advancement in several areas such as home to smart-home, cities to smart-cities, school to smart-school, health-care to smart-health-care,  and etc. The concept behind this ecosystem is the diversity of \emph{things} that outputs in a large-scale devices. Each connected device (physical or virtual) in the system, should be trackable and the generated information from the device can be retrievable by other users regardless of their locations \cite{lightweight}. Nevertheless, it is necessary that only authorized users can have access to the system and its resources. Otherwise, it may face several security concerns such as data modification, identity theft and information leakage. Moreover, security and privacy problems remain a demanding challenge in such a giant scale adoption of IoT because of the following reasons: (1) Mostly the communications between these IoT devices are wireless which make the system more susceptible to different attacks, i.e. message tampering, eavesdropping  and denial-of-service attacks like \emph{mirai} attack \cite{mirai} etc. (2) Devices from different company-makers have resource constraints limitation  such as processing power, battery and memory capacity that do not allow to deploy advanced security solutions.

Numerous solutions concerning security and privacy for IoT environments have been proposed that provide the mainstream security requirements i.e. confidentiality, integrity and authentication \cite{cia}. However, due to its heterogeneous nature and having low resource devices, existing solutions cannot fulfill the desired security requirements in the upcoming large-scale IoT paradigm. Even though some security based solutions are efficient and secure but are commonly based on centralized mechanisms. A known mechanism of PKI (Public Key Infrastructure) faces with scalability issues in case of million nodes.

Moving towards decentralized architectures, Block-chain (BC) technology has acquired an enormous attention in regard to tackling security, anonymity, traceability and centralization. Ethereum \cite{ethereum} a famous public blockchain project was introduced in 2014 that run smart-contracts for BC participants  to write and execute the application code in a distributed way. Basically, Blockchain is a distributed ledger technology where each operation such as read, create, update and delete, is recorded in the form of a transaction. Any unauthorized user accessing data or any operations on the existing data can, therefore, be detected. Furthermore, smart contracts are used to enforce the policies of access control on the existing stored data. A number of researches have shown the integration of BC technology in different IoT use-cases \cite{Ali2018}  \cite{dorri2017towards} \cite{platibart} \cite{marwan} \cite{Darwish2018} \cite{iot} \cite{Roulin} \cite{pharma}.

\subsection*{Problem Statement and Contribution:}
As from the current research proposals, it has been found that blockchain has become a promising technology to meet the future of IoT security and privacy requirements \cite{Christidis2016} \cite{Hizam2019}. Several Authors \cite{authzone} \cite{dorri2017towards} \cite{access} \cite{hardjono2016cloud} \cite{lightweight} put efforts in decentralized security mechanisms for upcoming large-scale IoT systems. But the limitation to all the approaches is that: there is no device-level trust that can prove any particular zone to external entities in case of supposing the communication to occur between different IoT networks. 
The contribution of this paper are as follows:
\begin{enumerate}
	\item Implement a custom \emph{Behavior Monitor} in IoT-Blockchain setup that can store \& monitor IoT devices data and classify its behavior (normal or malicious) to prevent attacks.
	
	\item Applying a filter on sensor-level that can stabilise output from single/multiple sensors to avoid faulty or malicious sensors in the network.
	\item To implement Trusted Execution Environment (TEE)) on a local blockchain of each IoT-Zone that ensure the integrity and confidentiality of sensitive application code and data.
\end{enumerate}

%\subsection*{Outline}
%Section \ref{bac} discussed background related to proposed architecture in detail. In Section \ref{rel} some past researches in the integration of Blockchain and IoT are mentioned. Section \ref{arc} discusses the proposed architecture and all its components in detail. In Section \ref{eval} we demonstrate evaluation results we have found while making experiments on IoT devices data. Finally Section \ref{conc} and \ref{future} we concludes the proposed work and discusses the future plan.

\section{Background}\label{bac}
\subsection{Internet of Things (IoT)}
The Internet of Things is the interconnection of smart-devices, mechanical and digital machines, objects and people which are capable of transferring data over the network without any human intervention. On the broader scale, IoT applications areas are smart-homes, smart-cities, smart-healthcare etc. The major components \cite{iotcom} in IoT ecosystem includes:
\begin{itemize}
	\item Smart-devices \& Sensors: The first layer is the device connectivity layer of IoT network, which constitutes different sensors like temperature sensor and thermostat, humidity sensor and many more.
	
	\item Connectivity: Devices in IoT are connected to low power wireless networks like LoRAWAN, ZigBee and Wifi etc. 
	\item Gateway: It acts as a middle layer between devices and manages the bi-directional transmission between networks and protocol. One of the key function of a gateway is to translate different protocols and make them interoperable.
	\item Cloud: This component integrates billion of sensors, smart-devices gateways, data storage and provides different predictive analytics.   
	\item Analytics: This is the process of converting the raw data (analog) of billion of devices into useful insights which can be further used for detailed analysis. 
\end{itemize} 
\subsection{Blockchain - a decentralized technology}
Blockchain technology was initially introduced and brought in 2008 and used by a remarkable known cryptocurrency, Bitcoin \cite{bitcoin}. It is a decentralized ledger technology that builds on a peer-to-peer network. Each node in the BC network holds an updated ledger copy that can hinder from a single point of failure. 
In the previous few years, the blockchain mostly based on cryptocurrencies such as  \cite{bitcoin} \cite{hyperfabric} in order to catch and addressed the double-spending problem. However, recently numerous other areas have been explored like IoT networks, where the blockchain can be set up to create and maintain digital transaction records in a secure and distributed fashion.

\begin{figure}[!h]
	\centering
	\includegraphics[width=\linewidth]{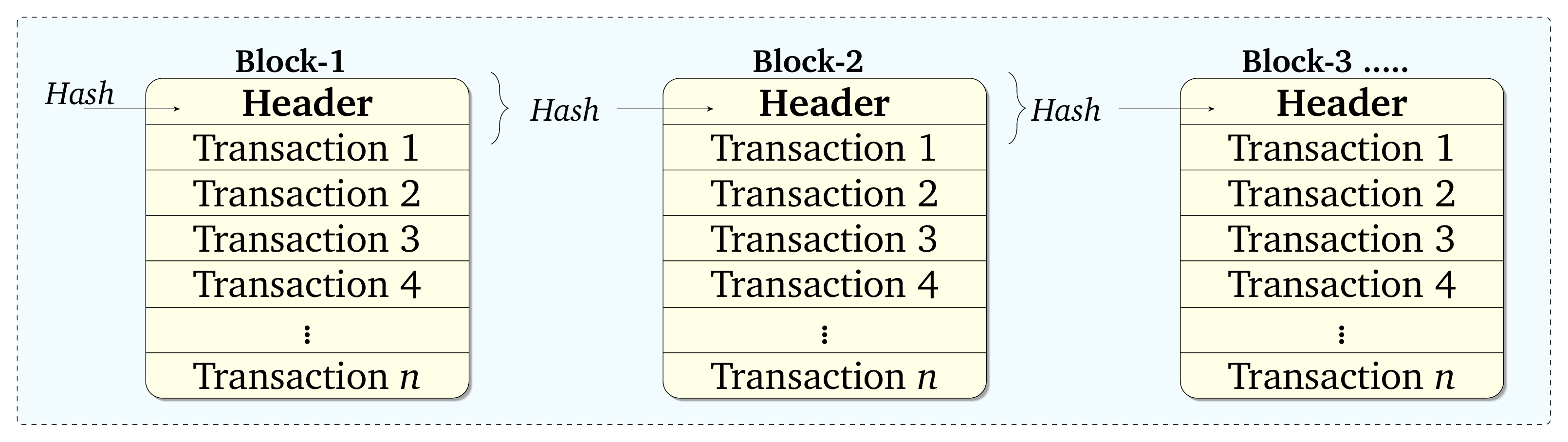}
	\caption{Inter-Linked Blocks in Blockchain}
	\label{bc}
\end{figure}

The ledger in BC is composed of blocks, and each block contains two parts. The first part represents the transaction (must be stored in a database), which can be of any kind, such as patient record, network traffic log, goods transaction etc. The second part includes the header detail such as hash of current transaction, concatenated previous hash and a timestamp. Thus, storage in this way makes a sequenced block of linked chain as shown in Figure \ref{bc}. Furthermore, if a new transaction starts, it will first add to certain block. Secondly, miners verify the block contain the transaction according to already defined rules. After verification process, a group of miners perform a consensus strategy to validate the transactions. Finally, upon successful validation the verified transaction is ready to append in the BC ledger.   
\subsection{Blockchain and IoT Systems}
IoT devices generate a large amount of data, that must be appropriately stored and analyzed for useful purposes. For each operation (create, update, delete, read) from IoT devices, the data can be treated in the form of transaction in the BC-blocks. Device identity information can be registered in a block such as manufacturer identity and the live-status of the device where it is located. Smart-contracts are used to enforce access control policies for IoT devices which can identify and detect unauthorized access. There is no need for a centralized authority for storage, such as cloud configuration, for IoT protection. Blockchain provides data authenticity, data integrity, traceability and prevents from unauthorized access. Blockchain technology can also enable a secure channel of messaging between IoT devices. Exchange of messages from one device to another device can be handle like financial transactions flow in crypto-currencies, such as ethereum \cite{ethereum} and Bitcoin \cite{bitcoin}.

\subsection{Blockchain Security Solutions for IoT}
The decentralized fashion of blockchain makes it a promising security solution for IoT paradigm. IoT and blockchain integration enables a higher and sound security level that could not be accomplished by any other technology or nearly impossible. Some of the recent research proposals in regards to IoT security with blockchain are as follows:

In \cite{managing}, authors proposed a blockchain-based approach for managing IoT devices and configurations. A unique paired-key (Public \& Private) is assigned to each device in the network. The private key is kept inside device and the public key is registered as a transaction in the blockchain. An IoT device can then be reached and access through ethernet by its public key. Hence, it is concluded that the management and control of IoT devices through blockchain is possible.

A study proposed in \cite{lee}, which make use of blockchain for secure firmware updates in IoT devices where traffic directly to the network server is replaced by local designated peers in the blockchain. The manufacturer is responsible to store the hashes of updated firmware on the blockchain peers that can be easily accessible to all the IoT nodes.

IoT devices using in medical and healthcare zone are also exposed to the same security and privacy limitations. In the case of health-care IoT system, it must be attack resistant and reliable enough. User safety and privacy is very critical and must be protected from any malfunction caused by a security incident or imprecise/faulty device. The risk of device malfunction can overcome in blockchain by immutable ledger technology. Nichol et al. \cite{medical} proposed the feasibility of BC in order to provide reliability in medical IoT devices. At the beginning when a device is manufactured and installed, a hash of UID (unique identifier) along with the other relevant information like manufacturer information, are stored in BC. Afterwards, this data will be updated with doctor-name, patient-history, and information about the hospital. The doctors and patients can be automatically informed about the device status like battery expiry, irregularities found in patient health. 
\subsection{Blockchain \& Trusted Execution Environment (TEEs)}
Trusted Execution Environments (TEEs) \cite{sgxonline}\cite{sgxonline1} have been utilized to enhance security and efficiency in the blockchain protocol. TEEs ensure confidentiality and integrity to the sensitive part of application code in the system, until and unless the CPU is not attempted physically by an attacker. TEEs also support remote attestation \cite{johnson2016intel}, that allows remote systems to verify the health of software with genuine TEE.

Intel provided TEEs functionality in \emph{Software Guard Extension} (SGX) \cite{sgxonline}. SGX is a set of CPU instructions inside Intel's x86 processor design which can allow creating an isolated environment for the execution of selected pieces of code in protected areas called \emph{enclaves}. These enclaves are designed to run software in a trustworthy environment, even on a system (host) where the operating system and memory are untrusted. There are three main functions of enclaves which are \emph{isolation}, \emph{sealing} and \emph{attestation}. A short description are as follows:
\begin{itemize}
	\item Isolation: Data and code inside the enclave part are protected and no access is allowed, such as read or alter by any external process.
	\item Sealing: Data that is supposed to send it to host environment should be encrypted and authenticated with a \emph{seal} key.
	\item Attestation: Remote systems or parties are allowed to verify an application enclave identity, credentials and other data.
\end{itemize}

%\subsection{Threat Model}

%The threat model for this research are as follows:
%\begin{itemize}
%	\item The proposed solution will only work on permissioned blockchain with TEE enabled systems.
%	\item Trusted execution environment (TEE) is considered as a root-of-trust so the TEE, CPU and hashing algorithm are considered trusted.
%	\item Network between IoT devices to the behavior monitor system is considered secure. No assumptions are made about software running alongside the behavior monitor. There can be any number of malwares trying to exploit the transactions by IoT devices in the internal network.
%\end{itemize}

\section{Related Work} \label{rel}
Currently, several types of research have been proposed in the integration of blockchain and IoT. A few of them have shown interest to help IoT security requirements. This section describes some of the past research proposals that intend to realize such integration, mainly for security needs. 

Raja et al.\cite{dorri2017towards} demonstrate blockchain-based architecture for smart-home setting. The architecture consists of three different blockchain networks: a local-BC (private), a share BC (private) and overlay BC (public). Although this research solves the issue of identification, still it has some shortcomings such as (1) For each operation it happened to make at least eight communication links that can flood the network quickly in case of high activity of IoT devices. (2) Local BC's are controlled by centralized entity which is opposite to the main principle of BC - a decentralized technology. 

In \cite{fairaccess}, authors study existing proposed models of access control systems and argue regarding these systems are not effective in the upcoming large-scale IoT. In order to avoid centralized mechanisms, this proposed research implements capability and access control as a sub-component in a blockchain environment. The other components are data management protocol, messaging service and data storage system. The messaging service deals with the exchange of access control message among two parties with defined roles. The messaging service then sends a request to the data storage system, where it is stored in the form of block. Finally, the receiving party fetch the message from the BC block using the messaging service. Moreover, they defined four roles, i.e. data owner, data source, requester and endorser. 

A mechanism named as \emph{chainanchor} proposed in \cite{hardjono2016cloud} based on the authorization of IoT devices in the cloud network. It helps device-owner being rewarded upon selling their device data to a service provider and ensure a privacy-preserving communication between owner and service-provider. But this approach is not suitable in most IoT use-cases, because the main scope of this research is full anonymity and IoT devices sometimes need device identification.

Patrick et al. \cite{authzone} introduced a decentralized authentication scheme for IoT devices. In this scheme they declare a separate virtual zones for each use-case such as healthcare zone, smart-school zone for robust identification of smart-devices. Each zone has a group master who is responsible to create a groupID and communicate with blockchain. Each device or follower in a zone gets a ticket signed by their respective zone master. When a device or follower wants to initiate a transaction,  an association request signed by private-key is send to their respective zone master. Upon receiving the request, BC verifies its integrity with the public key of follower. Afterwards, the follower ticket is verified using the master public key. If the ticket found valid, BC stores the association of followerID with their groupID for further correspondence, otherwise discarded. However, the limitation of this approach is that there is no mechanism that can provide trust-level confidence in each zone to prove it to the outside community.

To summarize, the majority of these current research proposals follows the same security schemes provided by existing BC technologies, i.e. Bitcoin \cite{bitcoin}, Ethereum \cite{ethereum} etc. However, there is no awareness towards device level trust that means to know the status of running IoT device, whether it is normal or malicious.

\begin{figure}[!th]
	\centering
	\includegraphics[width=\linewidth]{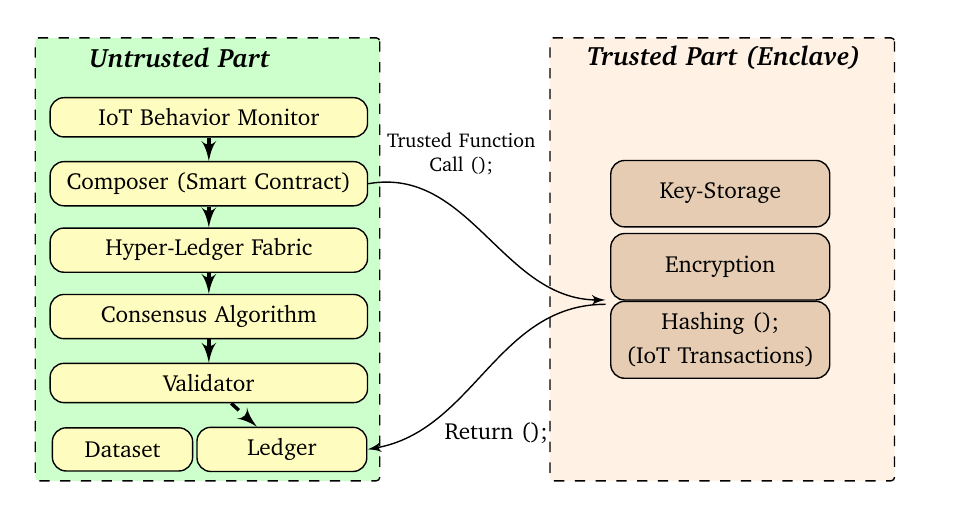}
	\caption{IoT secure behavior capturing and storage environment using TEE}
	\label{en}
\end{figure}

\section{Proposed Framework} \label{arc}
The main goal of the proposed framework (cf. Figure \ref{arch}) is to add and implement a security module for behavior monitoring on IoT-zones in a blockchain setup. As discussed in \cite{authzone}, authors declare zones for different use-cases of IoT. However, they do not consider the devices itself in case of infected behavior. Furthermore, there is no mechanism that can show the trust-Level confidence of each zone when an external entity needs to know before establishing connection. In this research, we extend the above scheme and add a behavior monitoring module on each zone. A separate local-BC is configured on each zone that is used to store the activity of each zone and provides the trust-level confidence to the other zones.
\begin{figure}[!th]
	\centering
	\includegraphics[width=\linewidth]{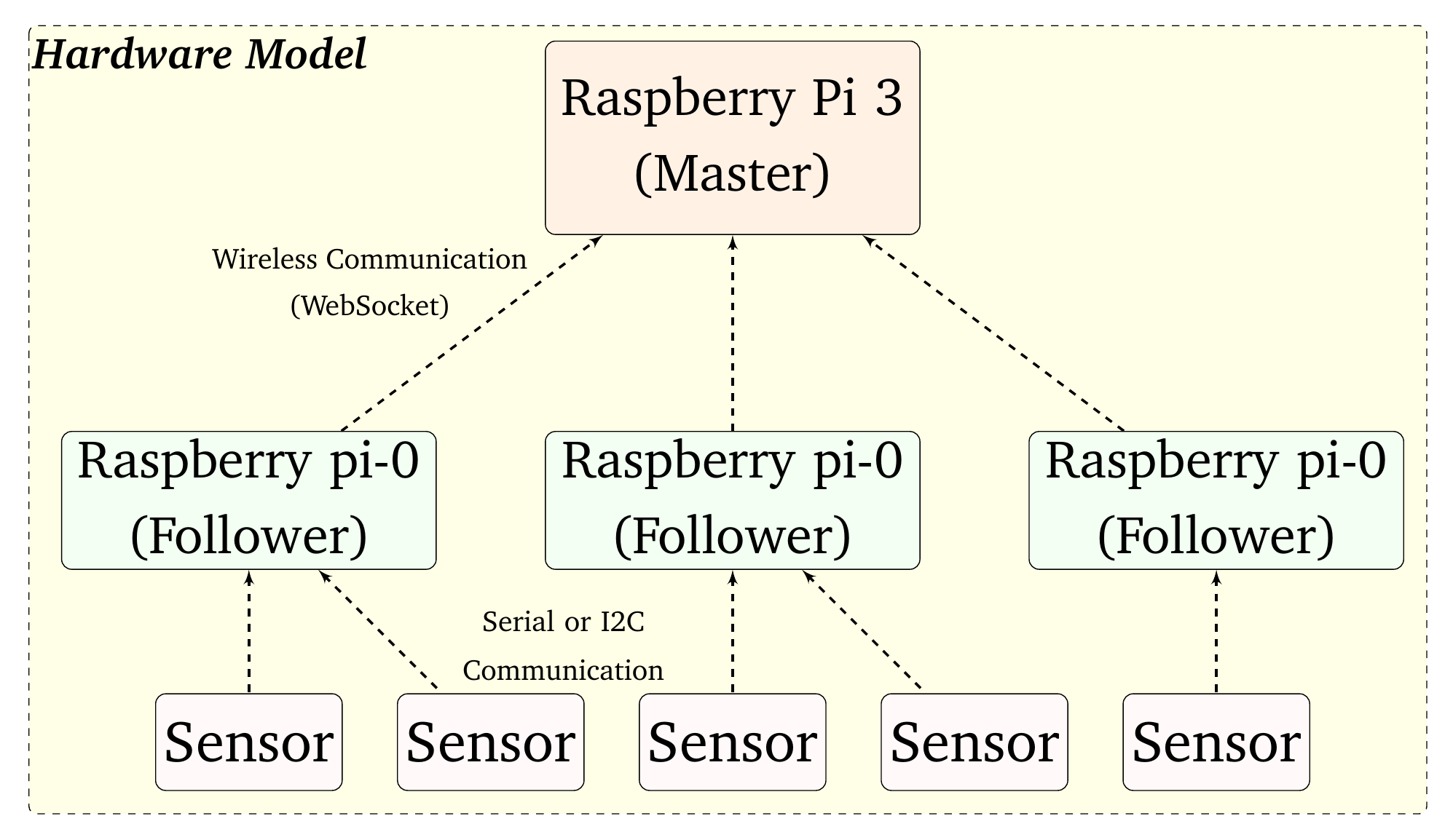}
	\caption{Hardware Model for IoT Zone} \label{hardware}
\end{figure}

All the communications passes betwen devices are considered as transactions and must be direct through the blockchain for validation. For example, if node \emph{A} need to send a message to node \emph{B}, then \emph{A} must first send the message to blockchain. If BC validates and authenticates the message from \emph{A}, then \emph{B} is finally allowed to read the message.
\subsection{Initialization \& System Functioning}
In the first phase of deployment, a single device from each zone is designated as a \emph{Main or Master} node, which can be considered as a certification authority (CA). Any node can be defined as a master, but in this case, we assigned to the node that is more resource capable and powerful. All the other nodes in each zone are known as \emph{follower}. Every Master node creates a \emph{groupID} and send a signed \emph{ticket} to each follower for identification. For the first transaction of any follower, it must require authentication. After that, an association of the follower and master are stored in the BC for future correspondence.
\subsubsection{Hardware Model of IoT}
The hardware architecture we use in our proposed framework for prototyping consists of multiple \emph{raspberry pi's}. The main/master node is configured on raspberry pi-3 for the sake of more resources. Followers or clients node work on raspberry pi-0 with a direct connection to sensors and other digital devices. Wifi is used for communication between master nodes, and follower communicates to their sensors using \emph{serial} or \emph{I2C} communication protocol as shown in Figure \ref{hardware}.

Every device is assigned by a key pair that consists of a public and private key. The private key is stored in follower (pi-0), while the corresponding public key is stored in their respective master node (pi-3). The connection between the follower and master node is established through WebSocket. Upon a connection request from follower to master, the follower must be required to send digital signature. Afterwards, master node should validate the digital signature in the blockchain before a secure WebSocket authorisation.   
 
\subsubsection{Improving Sensor Level Data Accuracy}
In order to improve sensor level security, the data acquisition procedure will use \emph{Kalman filter} to make a data model based on single/multiple sensor readings and covariance. For example, the position of a drone can be estimated in 3-axis based on GPS, but GPS alone cannot guarantee accurate altitude. Similarly, a Barometer data can drift based on different weather conditions at same altitude. Radar or Lidar will output the altitude value from the ground, but if an obstacle supposed to happen between the ground and radar the readings might become inaccurate. To avoid such discrepancies, Kalman filter uses data from all the 3 sensors GPS, barometer and radar/lidar, to predict the correct value (3D location) based on the covariance. This way if a faulty or malicious sensor found, the Kalman filter will automatically filter out the data from that sensor.

\begin{figure*}[!th]
	\centering
	\includegraphics[width=0.99\textwidth]{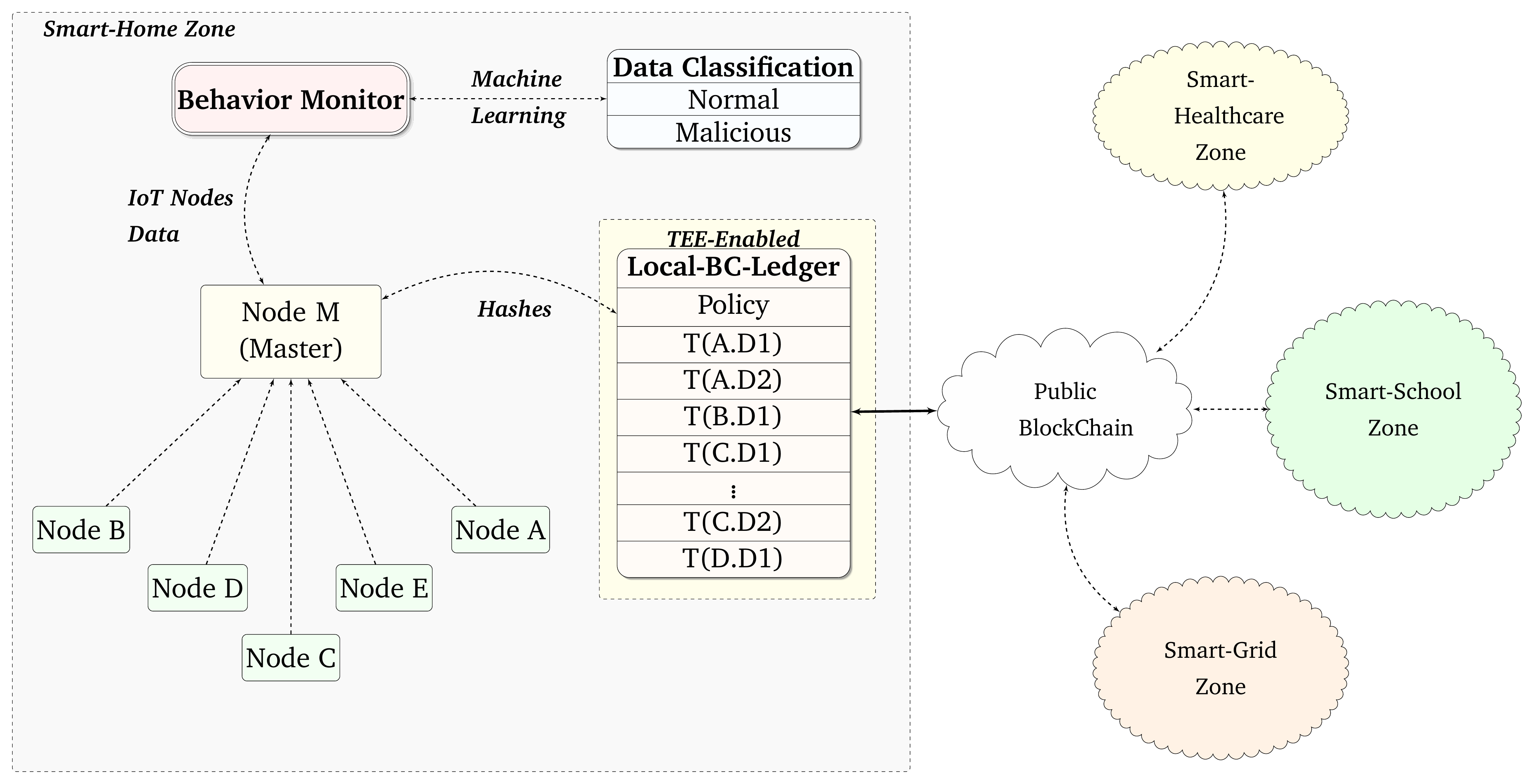}
	\caption{Proposed IoT Blockchain Framework} \label{arch}
\end{figure*}

\subsection{Configuring Local Blockchain}
A local private blockchain is deployed on a master node (Raspberry pi-3) of each zone and populated with the hashes of transactions generated from smart-devices. Hyperledger Fabric \cite{hyperfabric} a permissioned-BC is implemented as a local BC, we discussed the workflow of fabric with IoT in our previous research \cite{Ali2018}. 
For prototype implementation, we use the dataset \cite{dataset} of IoT traffic that has been collected from various sensor communication. For each communication between nodes or smart-devices, a transaction is created and stored in the local BC. Note that in majority of the current BC technologies, actual data of IoT devices are not stored in the BC due to overheads (i.e. processing \& network). 

In each zone, a single device having more computational power than others, acts as a \emph{master} or \emph{main} node. Likewise in our model we use raspberry pi-3 which is computationally and energy-efficient act as a \emph{master/main} node. Once the number of transactions reaches a pre-define \emph{blocksize}, the master node creates a new block and append it to local BC. Afterwards, we realise \emph{Intel SGX} \cite{sgxonline} as a root-of-trust on top of BC to ensure that the execution of sensitive code and applications are in trusted mode. As shown in Figure \ref{en}, the TEE-enabled application is composed of trusted and untrusted part. For sensitive operations like encryption and hashing a trusted-function is called. The function returns, and the data inside the trusted part (enclave) remains in trusted memory and are not accessible to external entities. Moreover, implementing \emph{SGX} technology on blockchain allows the proposed scheme to:

\begin{itemize}
	\item Provide protection to the applications and data running on BC.
	\item Make sure about the application and data running on the BC is as expected and correct.
	\item Provide end-to-end privacy to the application result, that cannot accessible by others to inspect but the user.
	\item Ensure a BC-based validation by verifying the applications inside \emph{enclave} is neither tampered nor interrupted by any node in BC.
	\item Make sure about the validity of application and execution results, and not tampered or fabricated by any malicious node.
\end{itemize}  

\subsection{Behavior Monitor}
The main achievement in this research is to define a behavior monitor that can classify the behavior of the devices and compute a level-of-trust for each zone. As mentioned earlier, all the nodes (followers) in a specific zone do their operations (read, write) via the master/main node. The scheme shown in Figure \ref{arch} depicts our proposed solution with all the entities in detail.  Data or transactions generated from device is considered as a behavior parameter of that device. The master node is a device that centrally manage and organize all the incoming and outgoing transactions. 

When the data is reached to the master from the follower node, the master node stores it in behavior monitor and append the corresponding hash to the ledger in blockchain. A sequence-ID is assigned to every generated transaction from the nodes while storing in behavior monitor, and a Hash-ID is also attached to the corresponding hash in BC, for future reference. Finally, a deep learning strategy is used to actively monitor the incoming data and classify them as normal or malicious. 

For the purpose of behavioral analysis and detection, we rely on Auto-Encoders (AE) - a deep learning model \cite{hinton2006reducing} \cite{nauman2017deep} for IoT devices, which is trained from statistical based correlation features extracted from benign set of data. The process of behavior detection and monitoring consists of the following sequential stages. (1) Data collection (2) Feature extraction (3) Training model (4) Continuous Behavior Monitoring.

\subsubsection{Data Collection}
At this point, this research work use the dataset \cite{dataset} that has been collected from various sensors in the smart-home IoT network. To ensure in real-time that the training dataset is clean and not malicious, normal traffic from IoT devices are collected immediately after its joining to the IoT network.  
\subsubsection{Feature Extraction} 
Whenever data from IoT device arrives, a behavioral snapshot of the protocols and host associated to the data are stored in the \emph{behavior monitor}. The snapshot contains different parameters, i.e. source \& destination IP, MAC-address and port number, etc. We use the same set of features included in the dataset for real-time detection of malicious activities in IoT devices. For example, when an infected node in a zone spoof an IP, then the features aggregated from the source \& destination IP along with MAC-Address will immediately mark as a malicious node because of unseen activity produced from the respective spoofs IP.

%\begin{figure*}[!th]
%	%	\centering
%	\begin{minipage}[b]{.55\textwidth}
%		\includegraphics[width=6cm, height=4cm]{barchart.pdf}
%		\caption{Detection Accuracy comparison with other Algorithms}\label{det}
%	\end{minipage}\qquad
%	\begin{minipage}[b]{.45\textwidth}
%		\includegraphics[width=7cm,height=4cm]{bar.pdf}
%		\caption{Detection time (seconds) comparison with different Algorithms'}\label{time}
%	\end{minipage}
%\end{figure*}

\subsubsection{Training Model}
As our baseline model for behavior detection, we use auto-encoder that can build and maintain a learning model on all zone of IoT network. An auto-encoder is a type of artificial neural network (ANN), which is trained to re-structure the data after some compression. The compression ensures that the model would be able to learn meaningful concepts and the correlation between different set of features.
For training purposes, we use two sets of data which consists of only benign (normal) data. The first dataset is a \emph{training dataset} $(T_{DS})$ which is used to train the auto-encoder by declaring input parameters such as \emph{learning rate} $(lr_{n}$, size of gradient descent step), and $epochs$ (number of iterations through $T_{DS}$). The second dataset $Opt_{DS}$ \emph{(Optimization Dataset)} is used to optimize the above hyper-parameters ($lr_{n}$ \& $epochs$) iteratively until the mean square error $(MSE)$ function between the input and output stop decreasing. This stopping prevents overfitting in $T_{DS}$ and help out better detection results with future data. Later on, $(Opt_{DS})$ is used to identify normal and malicious activities and false positive rate (FPR).

%----------------------------------graphs

After completing model training and optimization, a threshold value $(th^v)$ is set by which an instance of data is considered malicious.
Empirically, it is calculated by the sum of sample mean along with the standard deviation of $MSE$ on $Opt_{DS}$ (see Equation \ref{eq:1}).

\begin{equation}\label{eq:1}
th^v = \overline{MSE}_Opt_{DS} + s (MSE_Opt_{DS})
\end{equation}
\begin{figure*}[!th]
	\centering
	\includegraphics[width=0.70\textwidth]{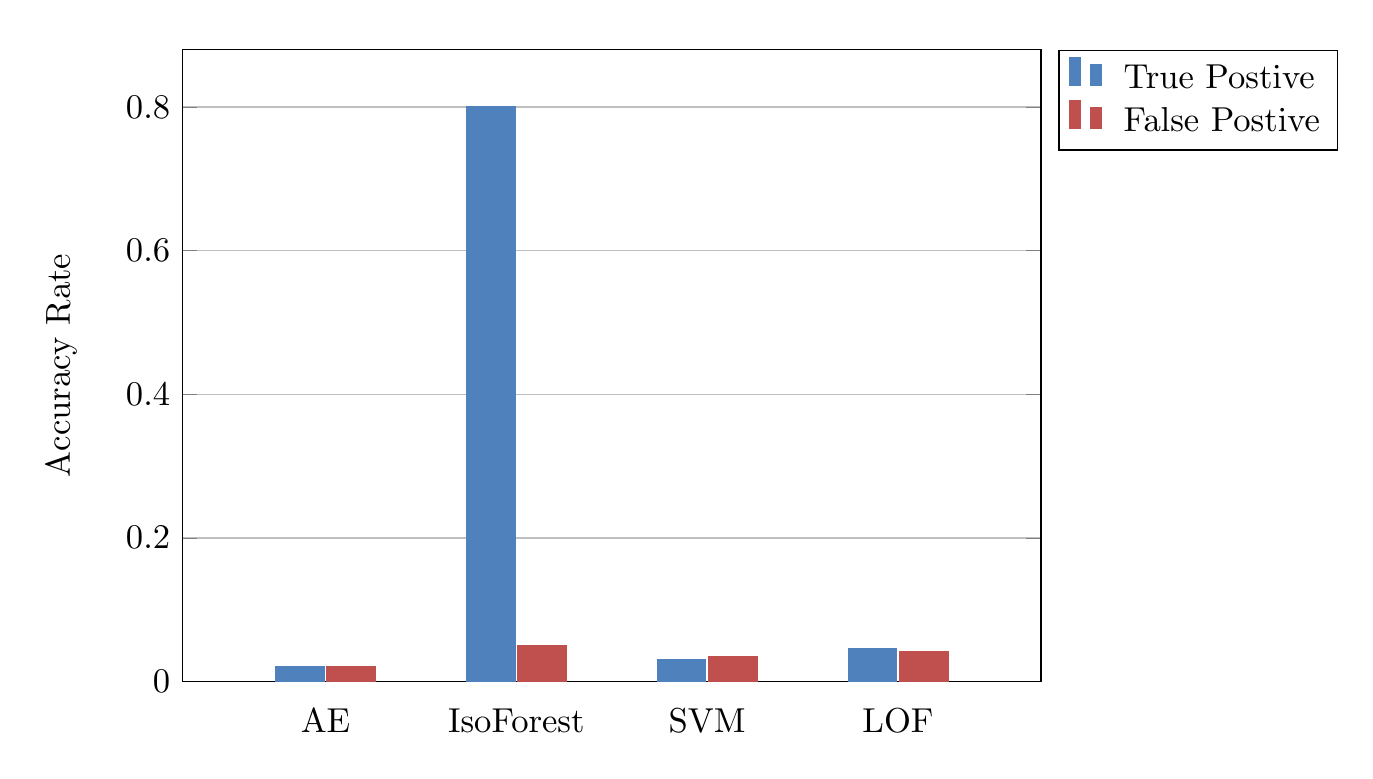}
	\caption{Detection Accuracy comparison with other Algorithms} \label{barchart}
\end{figure*}
\begin{figure*}[!th]
	\centering
	\includegraphics[width=0.75\textwidth]{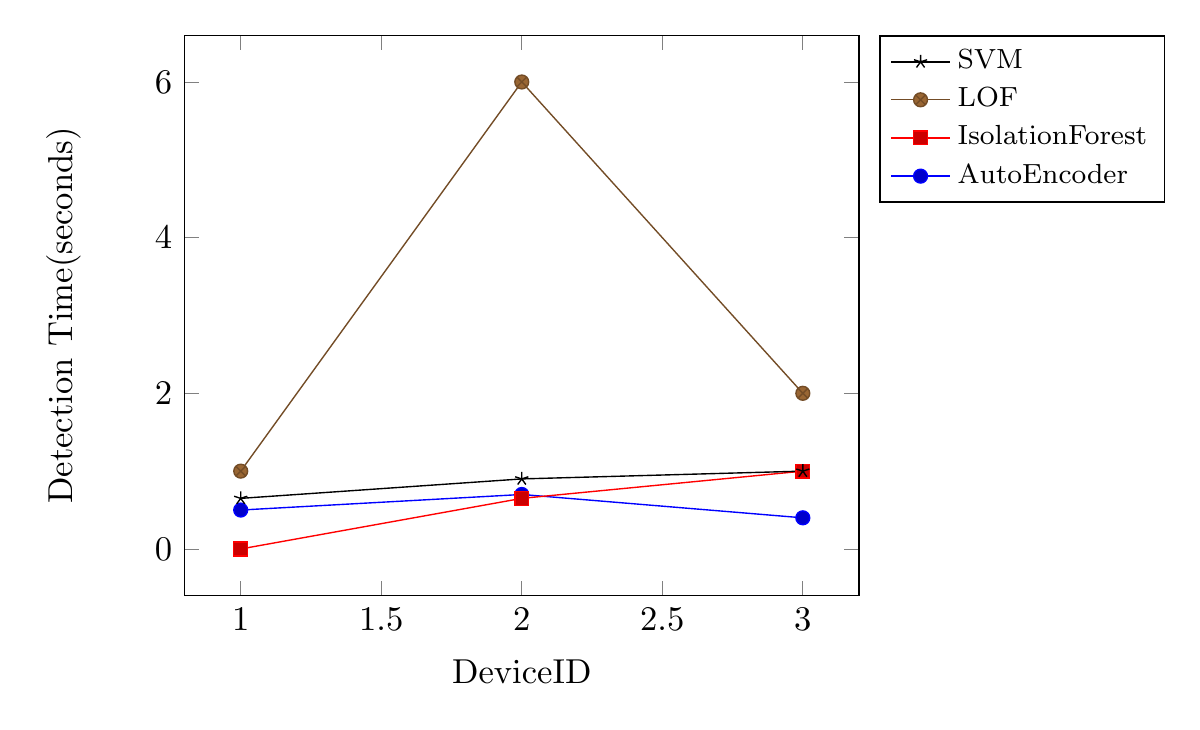}
	\caption{Detection time comparison with other Algorithms} \label{bar}
\end{figure*}
\subsubsection{Continuous Behavior Monitoring}
Finally, the model is applied to continuously observe the data and to label each instance as normal or malicious. Consequently, an alert against abnormal behavior can be issued to indicate the IoT device is malicious. Afterwards, for each IoT zone the behavior monitor calculates a trust-level measurement and a threshold must be defined for every use-case. Whenever a user or node from outside need to accessed data from any specific zone, our model is capable of disclosing the health of zone before establishing connection. This way a trusted environment can be built and informed the user about the state of any particular zone before actual communication.

\section{Experimental Analysis}\label{eval}
In our experiments, we use a real-time large dataset available in \cite{dataset}, for realizing the proposed framework. The dataset contains both benign and malicious (attacked) data. The data we choose from the dataset belongs to three different devices which are Ecobee-thermostat, Webcam and Security-camera. For training and optimization, we use \emph{tensorflow} and \emph{keras} libraries in python language. An auto-encoder make an input layer whose dimension is the same as the number of features in the dataset, i.e. 115. 

After training, we apply a famous DDOS attack known as (\emph{mirai}) to calculate the detection time and accuracy of our model in comparison with other machine learning algorithms. The same benign dataset is used to train three other algorithms: SVM (support vector machine), Isolation forest and LOF (\emph{Local Outlier Factor}). Our method shows 99.2\% results in terms of TPR (True Positive Rate) and fewer FPR (False Positive Rate). Furthermore, as evident in Figure \ref{barchart} SVM and LOF have almost similar TPR value and found much better than the isolation forest. 

Next, we evaluate the average detection time for each algorithm as depicted in Figure \ref{bar}. The detection time recorded for all the three devices is lower than the others in our case. The deep auto-encoder strategy outer-perform on all the selected devices in terms of False-positive, True-positive and detection time. This is because of the ability in auto-encoders to learn approximate complex functions and non-linear structure mapping \cite{detection}. 
Moreover, as shown in Figure \ref{bar}, our technique required much less time than the other algorithms which is approximately 175$\pm$230ms to detect the attacks. This means that the launch attack could be detected or alerted in less than a second and thus considers as a  substantial reduction in a typical time required for DDOS attacks \cite{DDOS}.

\section{Conclusions and Future Work}\label{conc}
In this research, we introduced a mechanism that can analyze  device level trust in IoT-Blockchain Infrastructure. A smart-home network is used as a use-case for realizing the proposed research idea. For prototype implementation a Local Blockchain on each zone is deployed on a master (raspberry pi-3) node that can store every traffic coming from their follower (Raspberry pi-0)) in the form of transactions. Behaviour Monitor is defined and configured on the Main/Master node of each zone, which is capable of capturing and analyzing the runtime activity of IoT devices. We apply a deep learning strategy (auto-encoders) for realisation on the behavior monitor to classify the device and make a level-of-trust.
Furthermore, we incorporate Trusted Execution technology (TEE) as a root-of-trust over the blockchain to provide security for sensitive code and applications. Finally, the proposed framework could meet the current security problems in IoT-Blockchain environment. And the evaluation of our study shows its ability to mitigate the mainstream security requirements and resilience to attacks.

This research work is our first step towards classification of devices in  IoT-Blockchain framework by means of deep learning. Our future plan is to investigate a comparative study of other machine learning approaches for better results in terms of performances and accuracy. Another goal would be to realize the framework in other use-cases of IoT domain and analyze the outcomes. Finally, in the near future we will provide a full implementation on various IoT devices datasets along with full verification mechanism of zones in a trusted way and make the source online to research community.

 \bibliographystyle{splncs04}
 \bibliography{sample-base}

\end{document}